\documentclass[english,preprint,superscriptaddress,a4paper]{revtex4-1}
\usepackage[T1]{fontenc}
\usepackage[utf8]{inputenc}
\usepackage{color}
\usepackage{babel}
\usepackage{amstext}
\usepackage{graphicx}
\usepackage{microtype}
\usepackage[unicode=true,
 bookmarks=true,bookmarksnumbered=true,bookmarksopen=true,bookmarksopenlevel=2,
 breaklinks=true,pdfborder={0 0 0},pdfborderstyle={},backref=false,colorlinks=true]
 {hyperref}
\hypersetup{pdftitle={SOL QCF},
 pdfauthor={RH},
 allcolors=blue}

\makeatletter
\usepackage{libertine}
\usepackage[libertine]{newtxmath}

\makeatother

\begin{document}
\title[SOL QCF]{Observation of quasi-coherent density fluctuation in scrape-off layer
enhancing boundary transport in high-$\beta_{\mathrm{N}}$ hybrid
plasmas on DIII-D }
\author{R.~Hong}
\affiliation{Physics and Astronomy Department, University of California, Los Angeles,
CA 90095, USA}
\author{T.~L.~Rhodes}
\affiliation{Physics and Astronomy Department, University of California, Los Angeles,
CA 90095, USA}
\author{Z.-Y.~Li}
\affiliation{Oak Ridge Associated Universities, Oak Ridge, TN 37831, USA}
\author{H.~Wang}
\affiliation{General Atomics, San Diego, CA 92121, USA}
\author{L.~Zeng}
\affiliation{Physics and Astronomy Department, University of California, Los Angeles,
CA 90095, USA}
\author{K.~Barada}
\affiliation{Physics and Astronomy Department, University of California, Los Angeles,
CA 90095, USA}
\author{G.~Wang}
\affiliation{Physics and Astronomy Department, University of California, Los Angeles,
CA 90095, USA}
\author{J.~G.~Watkins}
\affiliation{Sandia National Laboratories, Post Office Box 969, Livermore, CA 94551,
USA}
\author{W.~A.~Peebles}
\affiliation{Physics and Astronomy Department, University of California, Los Angeles,
CA 90095, USA}
\begin{abstract}
We report the observation of a quasi-coherent density fluctuation
(QCF) by the Doppler backscattering system in the scrape-off layer
(SOL) region of the DIII-D tokamak. This QCF is observed in high power,
high performance hybrid plasmas with near double-null divertor (DND)
shape during the electron cyclotron heating period. This mode is correlated
with a steepened SOL density profile, and leads to significantly elevated
particle and heat fluxes between ELMs. The SOL QCF is a long-wavelength
ion-scale fluctuation ($k_{\theta}\rho_{s}\approx0.2-0.4$ and $k_{r}\rho_{s}\approx0.03$),
and propagates in the ion diamagnetic direction in the plasma frame.
Its radial expanse is about 1.5--2~cm, well beyond the typical width
of heat flux $\lambda_{q}$ on DIII-D. Also, the SOL QCF does not
show any clear dependence on the effective SOL collisionality, and
thus may raise issues on the control of plasma-material interactions
in low collisionality plasmas in which the blob-induced transport
is reduced. A linear simulation using BOUT++ with a 5-field reduced
model is performed and compared with experimental observations. In
simulation results, an interchange-like density perturbation can be
driven by the SOL density gradient, and its peak location and the
radial width of the density perturbation are in agreement with the
experimental observations. 
\end{abstract}
\maketitle

\section{Introduction}

The edge and scrape-off layer (SOL) turbulence plays an important
role in tokamaks, as it can affect the boundary plasma profiles, which
in turn set the boundary conditions for the core plasma, and thus
determines the overall confinement properties of the discharge \citep{hughes2002observations,osborne1998hmodepedestal,pitcher1997therelation,suttrop1997identification}.
The SOL turbulent transport can influence the particle and heat load
distributions on the plasma-facing components \citep{loarte2007chapter}.
These processes have strong impacts on the impurity concentrations
and neutral pressures in the peripheral region, which will also affect
the confinement and stability of plasmas.

As observed in previous experiments \citep{boedo2003transport,endler1995measurements,labombard2001particle,labombard2000crossfield,antar2008turbulence},
the radial turbulent transport in the SOL region is thought to be
dominated by intermittent filamentary structures, such as blobs \citep{krasheninnikov2001onscrape}.
The blobs and their effects on the SOL transport have been investigated
in L-mode and between ELMs H-mode plasmas \citep{zweben2015edgeand,zweben2016blobstructure,boedo2009edgeturbulence,fuchert2014blobproperties,ayed2009interelm,boedo2001transport,rudakov2002fluctuationdriven,garcia2007fluctuations},
to validate edge transport modeling of the SOL for future fusion devices,
such as ITER. The statistical properties and intermittent features
of the SOL turbulence are found to be similar in both \emph{L}-plasmas
and moderately powered \emph{H}-plamsas \citep{antar2008turbulence,rudakov2002fluctuationdriven}.
Also, the blob-induced transport increases as the density or the SOL
collisionality is raised \citep{labombard2001particle,carralero2015experimental,carralero2014anexperimental}.

When the tokamak is operated in high-power, high-performance scenarios,
it is unclear whether the blobs are still the only dominant turbulent
structure in the SOL region, because the boundary plasma profiles
behave differently from those with low-to-moderate power injection.
For instance, a shelf-like density profile is observed to span the
separatrix, and leads to a steepened density gradient in the SOL region
of hybrid plasmas with near-double null divertor (DND) configuration
during high-power, high-$\beta_{\mathrm{N}}$ operation ($P_{\mathrm{in}}$
up to 15 MW, $H_{98}=1.4\text{–}1.8$ and $\beta_{\mathrm{N}}=3\text{–}4$)
\citep{petrie2017improved,turco2015thehighbetan}. Such steepened
boundary profiles may be able to drive SOL fluctuations, other than
the blobs, that are large enough to influence the transport process.
The studies on the edge and SOL turbulent transport in these high-power
hybrid plasmas are still limited. Thus, the identification of the
dominant SOL turbulence structures will improve our understanding,
and hence catalyze the optimization, of the high-power, high-$\beta_{\mathrm{N}}$
hybrid scenarios. 

For many years, SOL turbulence has been extensively investigated in
both toroidal and linear fusion devices by the means of Langmuir probes
\citep{antar2008turbulence,boedo2003transport,labombard2001particle,labombard2000crossfield}.
However, the use of probes in high power operations are limited, as
it can be easily melted by the high heat flux and leads to increased
impurity contamination. In this study, the SOL turbulence are mainly
investigated by the Doppler backscattering (DBS) systems \citep{peebles2010anovel}.
Based on the DBS measurements, we report the experimental observation
of a quasi-coherent density fluctuation (QCF) in the SOL region at
the low-field-side (LFS) midplane during the high-power hybrid operation.
This SOL QCF is a long-lived, ion-scale fluctuation between ELMs during
the electron cyclotron heating, which is distinct from the intermittent
and `bursty' events such as blobs. The SOL QCF is highly correlated
with substantially elevated particle and heat fluxes into the divertors,
and likely driven by the steepened SOL pressure gradient due to the
shelf-like density profile across the separatrix. Also, the SOL QCF
does \emph{not} depend on the effective SOL collisionality, and thus
may be an issue for the control of plasma-material interactions in
low collisionality plasmas in which the blob-induced transport is
expected to be reduced.

This paper is organized as follows. Sec.~\ref{sec:Experimental-Setup}
describes the experimental arrangement and major diagnostics used
in this study. In Sec.~\ref{sec:Results} we describe the general
patterns and characteristics of the SOL QCF measured by DBS systems
and examine the influence of boundary profiles on the SOL QCF. Results
from a linear simulation using BOUT++ are also shown in this section.
In Sec.~\ref{sec:Discussion}, we briefly compare the experimental
observations and the linear simulation results. Some different features
of the SOL QCF from the blobs, as well as the implications to the
boundary transport and PMI studies, are also discussed. Section \ref{sec:Summary}
summarizes the major findings and conclusions. 

\section{Experimental Arrangement \label{sec:Experimental-Setup}}

\begin{figure}
\includegraphics[width=2.5in]{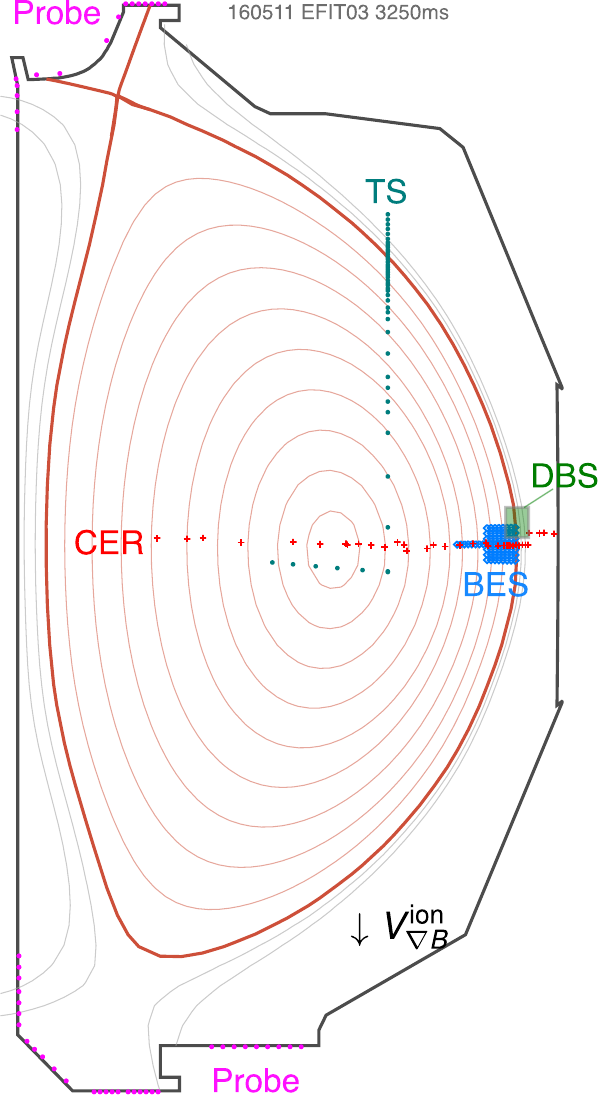}

\caption{\label{fig:diagnostics} Cross-sectional shape of discharges studied.
Positions of major diagnostics are shown: Thomson scattering (teal),
BES (blue), CER spectroscopy (red), Langmuir probe (magenta), and
DBS (green).}
\end{figure}

In this study, all discharges use the near double-null-divertor (DND)
configuration (Fig.~\ref{fig:diagnostics}) \citep{ferron2005optimization,petrie2017improved}.
These plasmas are featured by a slight magnetic bias toward the upper
(`primary') divertor, i.e., $dR_{\mathrm{sep}}$ ranges from $+0.5$
to $+0.7$ cm, where $dR_{\mathrm{sep}}$ is the radial separation
between the primary and secondary separatrix at the outer midplane.
The ion $\nabla B$ drift direction is away from the primary (upper)
divertor.

These high-performance hybrid plasmas are characterized by high energy
confinement ($H_{98}=1.4\text{–}1.8$) and high-$\beta_{\mathrm{N}}$
(3--4). Representative plasma parameters are: plasma current $I_{p}=1.0$
MA, the toroidal field $B_{T}=2.0$ T, the edge safety factor $q_{95}=6$,
and total power input $P_{\mathrm{in}}=11\text{–}15$ MW. The neutral
beam injection (NBI) provides up to 11.5~MW power input. The electron
cyclotron heating (ECH) system is configured for on-axis heating and
current drive with up to 3.5~MW. Line-averaged density was constrained
to no more than $5.5\times10^{19}\,\mathrm{m^{-3}}$ (about 60\% of
the Greenwald limit) during the ECH to avoid cutoff or reflection.
Also, argon seeding has been employed in some discharges, but it does
\emph{not} seem to change the behavior of the quasi-coherent density
fluctuation in the SOL.

The core and pedestal electron density and temperature in the mainchamber
(upstream) are measured by the Thomson scattering system \citep{eldon2012initial}
(teal in Fig.~\ref{fig:diagnostics}), while divertor electron density
and temperature (downstream) are measured by fixed Langmuir probes
(magenta in Fig.~\ref{fig:diagnostics}) \citep{watkins2008highheat}.
The charge exchange recombination (CER) spectroscopy \citep{chrystal2016improved}
provides carbon impurity temperature and density, as well as toroidal
and poloidal rotation speeds, at points along the outboard midplane
(red in Fig.~\ref{fig:diagnostics}). These impurity ion profiles
can be used to derive the radial electric field profiles using the
carbon ion force balance equation. A 2D array of high sensitivity
beam emission spectroscopy (BES) \citep{mckee2010widefield} is deployed
to measure the low-$k$ (e.g., $k_{\theta}<3\,\mathrm{cm^{-1}}$)
density fluctuations at the pedestal region (blue in Fig.~\ref{fig:diagnostics}).
An 8-channel Doppler backscattering (DBS) system is used to measure
turbulent flows and low-to-intermediate-$k$ density fluctuations
\citep{peebles2010anovel}. The cutoff layer and the wavenumbers of
the scattered positions of the millimeter waves at 8 different frequencies
emitted by the DBS system are calculated using the 3D ray-tracing
code GENRAY \citep{smirnov1994ageneral}. In this study, DBS measurements
cover the region across the separatrix, i.e., $0.9<\psi_{n}<1.1$,
where $\psi_{n}$ is the normalized poloidal magnetic flux surface
function (green in Fig.~\ref{fig:diagnostics}).

\section{Results\label{sec:Results}}

\subsection{Basic characteristics of SOL QCF\label{subsec:sol-qcf}}

\begin{figure}
\includegraphics[width=3.2in]{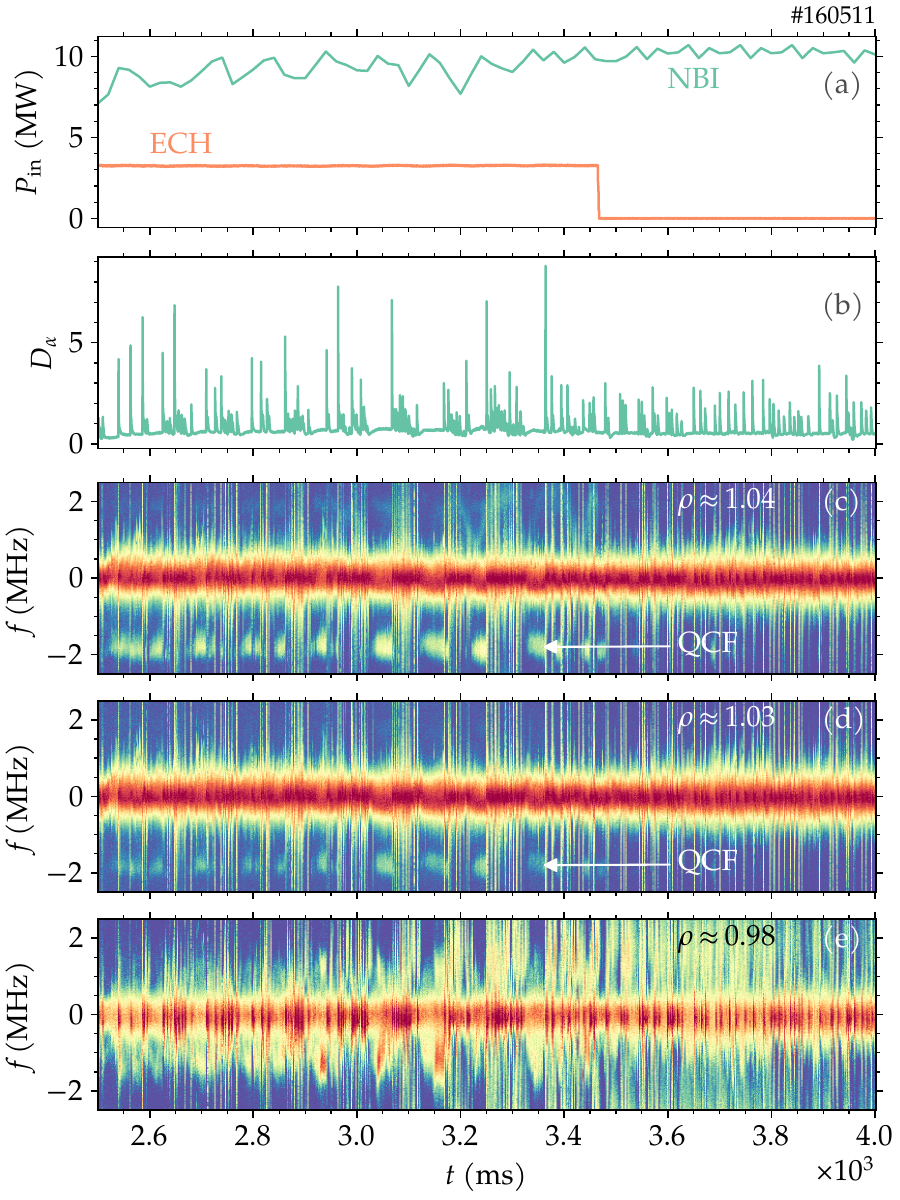}

\caption{\label{fig:sol-qcf-spectrogram} Time traces of (a) input power by
NBI (green) and ECH (orange), and (b) $D_{\alpha}$ line emission
intensity. Contour plots are spectrograms of the density fluctuations
measured by the DBS system at both SOL (c)--(d) and pedestal (e)
region.}
\end{figure}

We first explore some basic features of this quasi-coherent density
fluctuation, such as mode structures, phase velocities, and some statistical
properties. The SOL quasi-coherent fluctuation (QCF) is observed between
the ELMs when ECH is applied in addition to the 10 MW neutral beam
injection (Fig.~\ref{fig:sol-qcf-spectrogram}(a)). The burst of
the SOL QCF can be resolved spectrally and temporally by the DBS system
(Fig.~\ref{fig:sol-qcf-spectrogram}(c)--(d)). The cutoff locations
of first three channels are in the SOL region, and other channels
located at the pedestal ($0.9<\rho\leq1.0$). The frequency of such
density fluctuation spans from $-1.5$~MHz through $-2.2$~MHz with
$\Delta f/\bar{f}\approx0.3$. The poloidal wave-number of the QCF
is calculated by the 3D ray-tracing code GENRAY, which is found to
be $k_{\theta}\approx4\text{–}6\,\mathrm{cm^{-1}}$ or $k_{\theta}\rho_{s}\sim0.2\text{–}0.4$,
where $\rho_{s}=c_{s}/\omega_{ci}$ is the hybrid ion gyro-radius,
$c_{s}$ is the sound speed and $\omega_{ci}$ is the ion gyro-frequency.
Therefore, the SOL QCF is a long-wavelength ion-scale fluctuation.
The SOL QCF was not observed by the magnetic probe array. One reason
might be that the magnetic fluctuations in the $k_{\theta}$-range
of the SOL QCF ($k_{\theta}\approx4\text{–}6\,\mathrm{cm^{-1}}$)
is too weak to be detected by the magnetic probe array on the vessel
wall due to the quick decay of the high mode number fluctuations.

\begin{figure}
\includegraphics[width=3.3in]{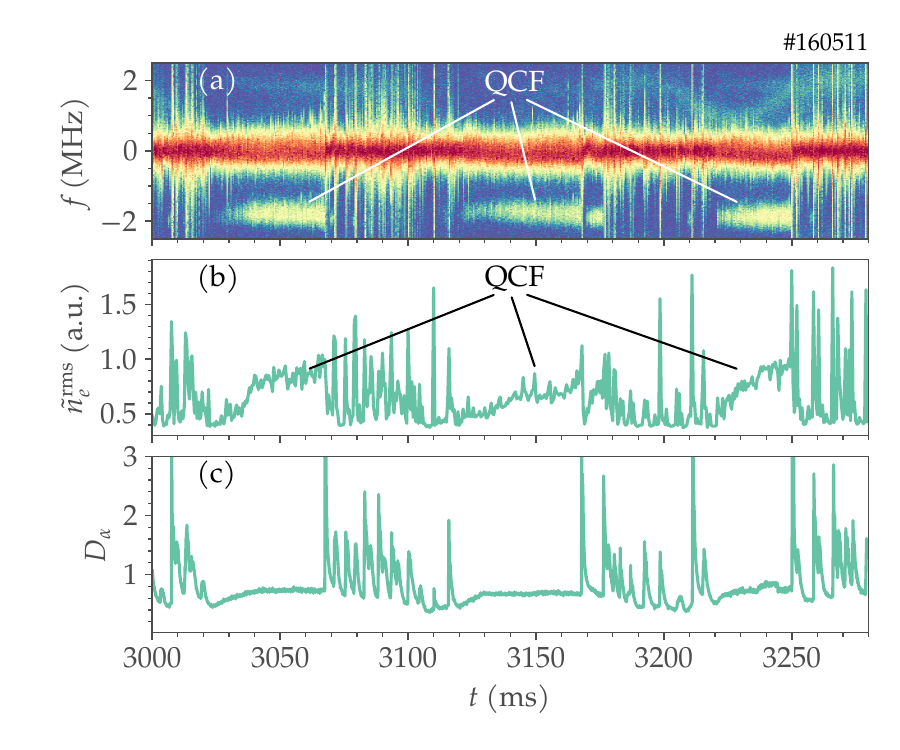}

\caption{\label{fig:qcf-between-elms}Spectrogram of SOL density fluctuation
signals in between ELMs (a), the RMS level of the SOL density fluctuation
at $-1<f<-2.5$~MHz (b), as $D_{\alpha}$ line intensity near outboard
divertor strike point (c).}
\end{figure}

The SOL QCF occurs rapidly after each ELM crash and substantially
increases the density fluctuation level. The evolution of the SOL
density fluctuation between a few ELMs in shot 160511 are demonstrated
in Fig.~\ref{fig:qcf-between-elms}. In this case, the SOL mode starts
to grow within 10~ms after each large type I ELM and the following
small ELMs (Fig.~\ref{fig:qcf-between-elms}(a)). The RMS level of
the density fluctuations corresponding to the SOL QCF start to saturate
in about 10--20~ms (Fig.~\ref{fig:qcf-between-elms}(b)). Also,
an elevated $D_{\alpha}$ line intensity are observed during the burst
of the SOL QCF (Fig.~\ref{fig:qcf-between-elms}(c)), implying a
potential role of the SOL QCF in the enhanced divertor transport and
recycling process.

\begin{figure}
\includegraphics[width=3.3in]{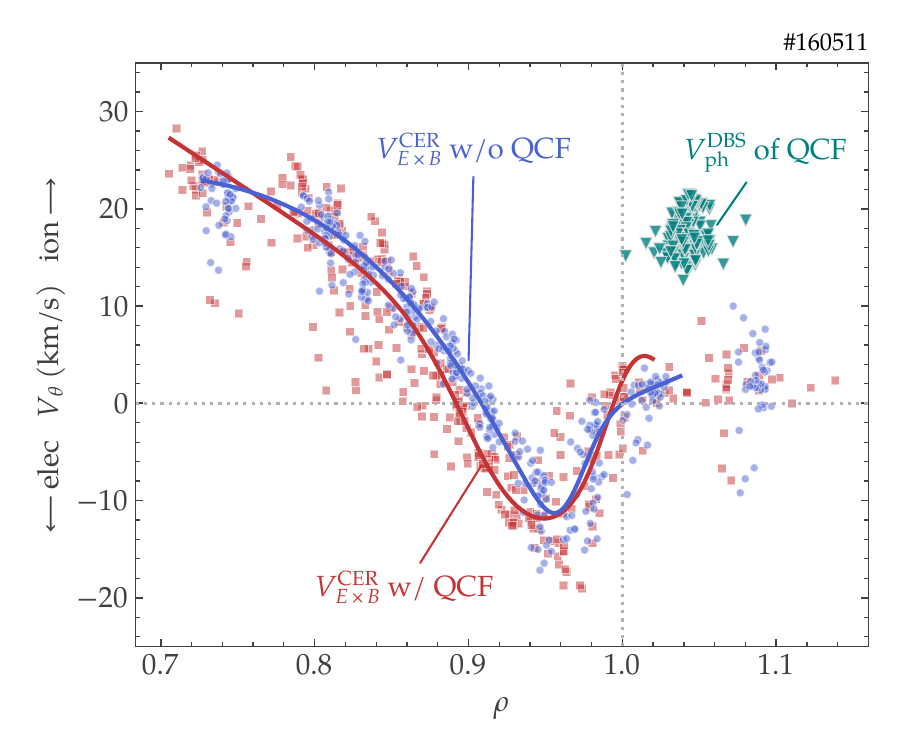}

\caption{\label{fig:Vpol} The radial profiles of poloidal phase velocities
of SOL QCF measured by DBS (teal triangles), and $E_{r}\times B$
velocities given by CER measurements with (red squares) and without
(blue circles) SOL QCF, respectively. Solid curves are the spline
fits of the scatter points. Positive velocity is in the ion diamagnetic
drift direction.}
\end{figure}

The phase velocity profile of the QCF can be estimated via $V_{ph}=2\pi\bar{f}/k_{\theta},$
where $\bar{f}=\sum_{f}fS(f)/\sum_{f}S(f)$ with $S(f)$ the power
spectrum density. The calculated phase velocity is a superposition
of the $E_{r}\times B$ velocity and any other poloidal velocity in
the plasma frame. The phase velocity of the SOL QCF is found to be
in the range of $-12$ to $-22$ km/s (teal triangles in Fig.~\ref{fig:Vpol}).
The positive velocity indicates the propagation in the ion diamagnetic
drift direction in the lab frame. The mean profiles of $E_{r}\times B$
velocities are also obtained at two different time periods, i.e.,
with (red squares in Fig.~\ref{fig:Vpol}) and without (blue circles
in Fig.~\ref{fig:Vpol}) detectable SOL QCF. Here, the radial electric
field, $E_{r}$, is obtained from the radial force balance equation
using the carbon impurity profiles measured by the CER system. The
mean $E_{r}\times B$ velocities do \emph{not} show significant differences
between those two cases. By comparing against the $E_{r}\times B$
velocities, we can see the SOL QCF is propagating in the ion diamagnetic
drift direction in the plasma frame.

\begin{figure}
\includegraphics[width=3.3in]{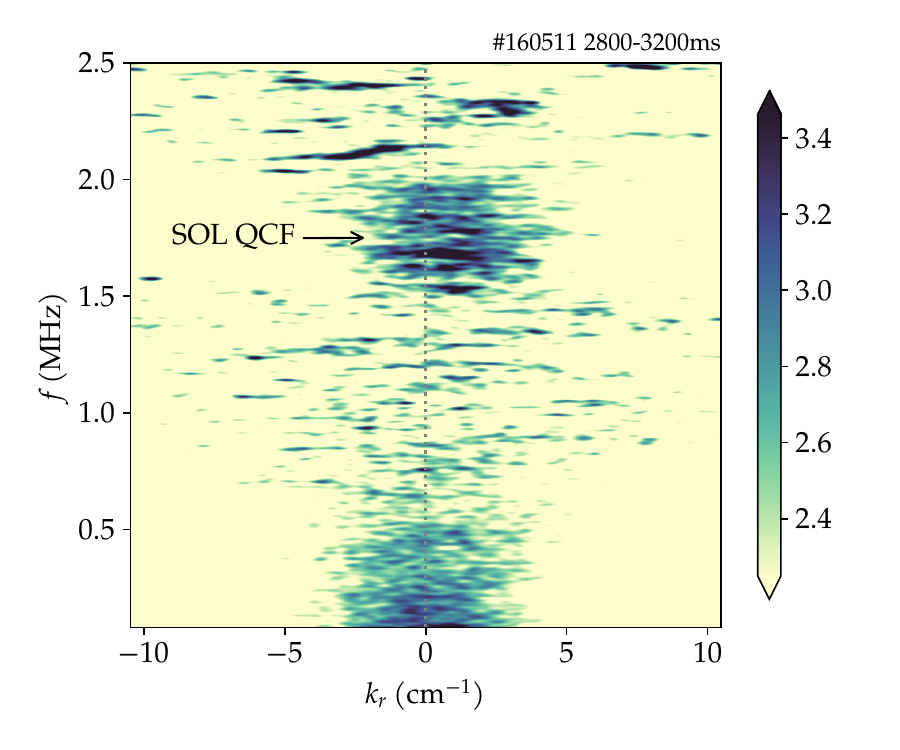}

\caption{\label{fig:kr} Wavenumber-frequency spectrum, $S(k_{r},f)$, in the
SOL region. The spectrum is calculated using the density fluctuations
measured by the channel 1 and 2 of the DBS system. Positive $k_{r}$
corresponds to the radially outward direction.}
\end{figure}

The radial wave-numbers of the SOL QCF can also be estimated using
the two-point technique \citep{beall1982estimation}. Figure~\ref{fig:kr}
shows the radial wavenumber-frequency spectrum, $S(k_{r},f)$, obtained
from two radially separated density fluctuations in the SOL region.
In $S(k_{r},f)$, the SOL QCF locates between 1.5~MHz and 2~MHz,
in contrast to the low-frequency components ($f<0.5$~MHz). The low-frequency
component may be related to the conventional filamentary events like
blobs. A positive value of the wave-number indicates a radially outward
phase velocity. The spectral-averaged radial wavenumber of the SOL
QCF is $\bar{k}_{r}=\sum_{k_{r}}k_{r}S(k_{r},f=f_{\mathrm{QCF}})/\sum_{k_{r}}S(k_{r},f=f_{\mathrm{QCF}})\approx0.5\,\mathrm{cm}^{-1}$,
where the $S(k_{r},f=f_{\mathrm{QCF}})$ is the wavenumber-frequency
spectrum in QCF's frequency range. In addition, the spread of the
spectral-averaged radial wavenumber is approximately $\Delta\bar{k}_{r}\approx3\text{–}4\,\mathrm{cm^{-1}}$,
where $(\Delta\bar{k})^{2}=\sum_{k_{r}}(k_{r}-\bar{k}_{r})^{2}S(k_{r},f=f_{\mathrm{QCF}})/\sum_{k_{r}}S(k_{r},f=f_{\mathrm{QCF}})$.
The radial correlation length of the SOL QCF can be inferred from
the spread of the radial wavenumber, i.e., $l_{c,r}=2\pi/(\Delta\bar{k}_{r})\approx1.5\text{–}2$
cm, which well exceeds the typical $\lambda_{q}$ mapped onto midplane
in the DIII-D (about 2--3~mm). 

\begin{figure}
\includegraphics[width=3.3in]{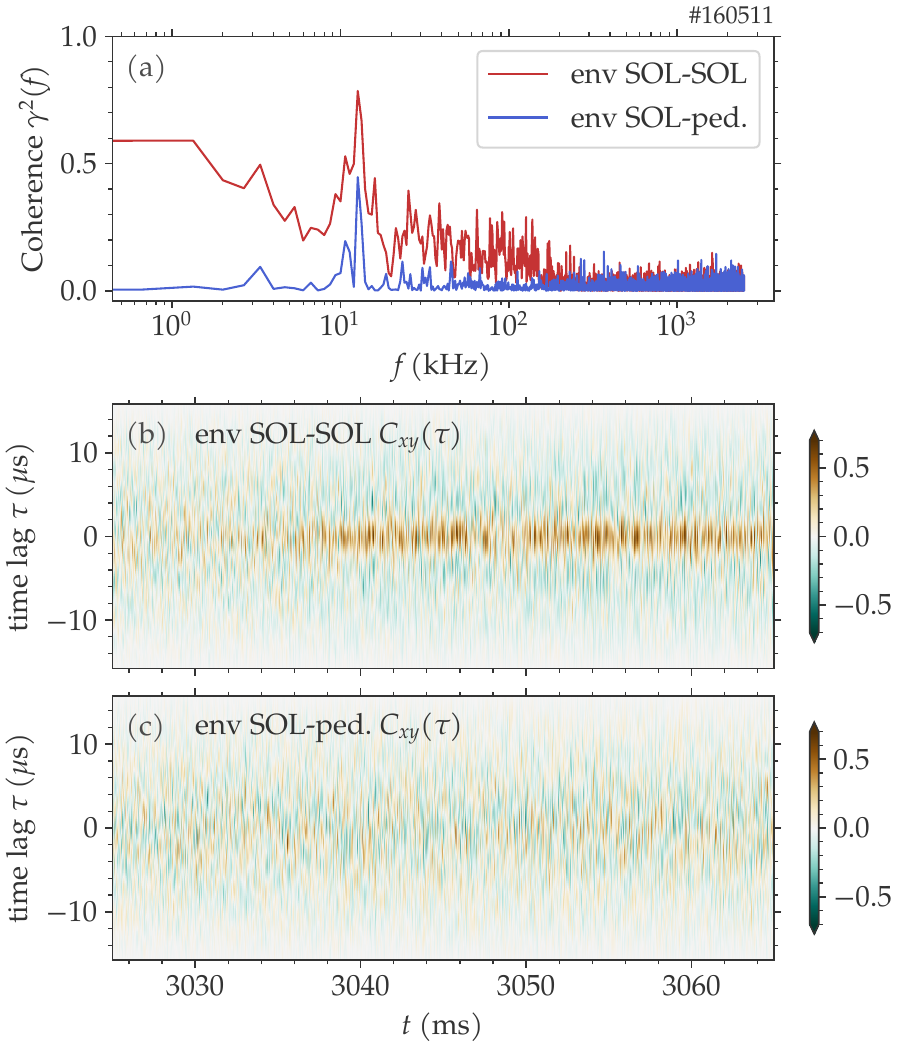}

\caption{\label{fig:coh-ped-sol}Coherence and cross-correlation analysis.
(a) Red curve: coherence between the envelopes of two density fluctuations
in SOL (channel-1 vs channel-2 of DBS); blue curve: coherence between
the envelopes of $\tilde{n}_{e}^{\mathrm{SOL}}$ and $\tilde{n}_{e}^{\mathrm{ped}}$.
(b) cross-correlation function between the envelopes of two density
fluctuations in SOL. (c) cross-correlation function between the envelopes
of $\tilde{n}_{e}^{\mathrm{SOL}}$ and $\tilde{n}_{e}^{\mathrm{ped}}$
($\rho\approx1.02$ and $0.98$).}
\end{figure}

The small radial wavenumber of the SOL QCF may raise a concern about
the link between the SOL and the pedestal fluctuations, i.e., whether
the SOL QCF is part of a long-range correlation structure extending
into the pedestal region. To answer this question, we compute cross-correlation
and coherence between the envelopes (fluctuation levels) of the density
fluctuations measured at the SOL and the pedestal region (as shown
in Fig.~\ref{fig:sol-qcf-spectrogram}(c) and (f)). The envelopes
are calculated by applying the Hilbert transform to the high-frequency
density fluctuations. The coherence between the envelopes of the SOL
($\tilde{n}_{e}^{\mathrm{SOL}}$) and the pedestal ($\tilde{n}_{e}^{\mathrm{ped}}$)
density fluctuation is weak (blue curve in Fig.~\ref{fig:coh-ped-sol}(a)),
except a peak at $f_{\mathrm{TM}}\approx13$ kHz. This frequency responses
to the $n=1$ tearing mode event that can modulate the profiles and
ambient turbulence. In contrast, the envelopes of two radially separated
SOL density fluctuations show larger coherence overall (red curve
in Fig.~\ref{fig:coh-ped-sol}(a)). Besides, the time-dependent cross-correlation
analysis shows that the phase delay between the envelopes of two SOL
fluctuations is close to zero (Fig.~\ref{fig:coh-ped-sol}(b)), while
there is no clear phase delay between the envelopes of $\tilde{n}_{e}^{\mathrm{SOL}}$
and $\tilde{n}_{e}^{\mathrm{ped}}$ (Fig.~\ref{fig:coh-ped-sol}(c)).
Note that in DIII-D the typical long-range correlation structures
usually show substantial coherence of envelopes even when $\Delta r/a\gg0.1$.
Thus, this weak coherence and cross-correlation, with relatively small
radial separation $\Delta r/a\approx0.04$, suggests that the SOL
QCF is unlikely a radial extension of the pedestal fluctuations.

It is worth noting that the separation distances used for those two
coherence calculations are different. The distance used in SOL-SOL
envelope coherence calculation is about $\Delta r\approx0.8$ cm,
and the one used in SOL-pedestal envelope coherence calculation is
about $\Delta r\approx2$ cm. However, the normalized distances in
terms of the local hybrid ion gyroradius, $\rho_{s}$, are similar,
since $\rho_{s}^{\mathrm{SOL}}/\rho_{s}^{\mathrm{ped}}\approx2-4$
in this case. Thus, it should be valid to compare those two coherence
calculation, though the radial separations used in the calculations
are not the same.

\begin{figure}
\includegraphics[width=3.3in]{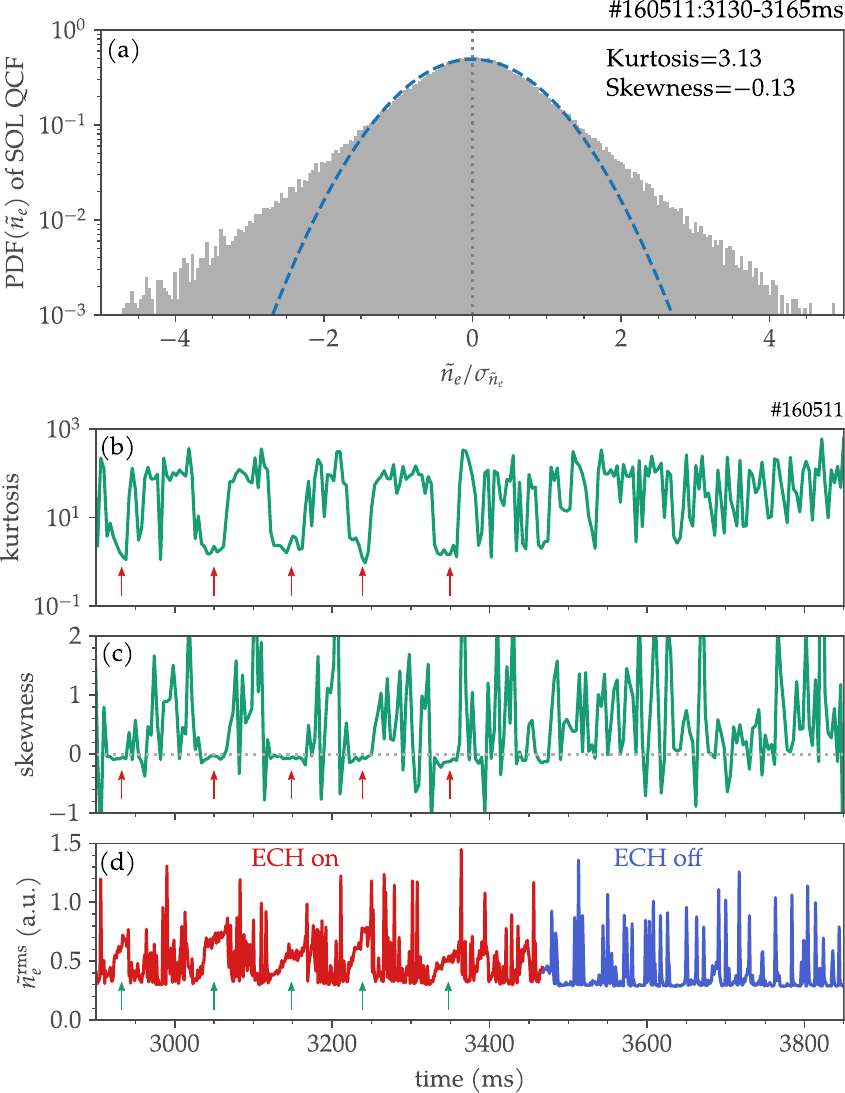}

\caption{\label{fig:sol-pdf} (a) The probability distribution density of the
quasi-coherent density fluctuation in the SOL region (gray bars) and
the Gaussian fit (blue dashed line) for the distribution. Time traces
of the kurtosis (b), skewness (c) and rms level (d) of the density
fluctuations in the SOL. The arrows in (b)--(d) indicate the periods
with the SOL QCF.}
\end{figure}

The statistical properties of the SOL QCF are also calculated using
DBS density fluctuation signals, but there is no clear features of
intermittency. In a typical discharge, the probability density function
(PDF) of the SOL QCF is heavier tailed (gray bars in Fig.~\ref{fig:sol-pdf}(a)),
i.e., shows fatter tails than the Gaussian distribution (blue dashed
line in Fig.~\ref{fig:sol-pdf}(a)). Here, the corresponding kurtosis
of PDF is $K=3.1$, and the Fisher's kurtosis definition is used so
that the standard normal distribution has a kurtosis of zero. Also,
the PDF is approximately symmetrical and thus leads to a negligible
skewness $S=-0.13$. The skewness of the SOL QCF is much smaller than
that of the blobs in the DIII-D which typically approaches a value
of one in the far SOL \citep{boedo2003transport}. The time traces
of the kurtosis and skewness are shown in Fig.~\ref{fig:sol-pdf}(b)
and (c). During the burst of the SOL QCF, the kurtosis of the SOL
density fluctuations drops to $1<K<5$ and the skewness drops to a
value about zero. When there is no SOL QCF observed, the kurtosis
of the SOL density fluctuations is typically at higher values, i.e.,
$10<K<100$, and the skewness approaches a value of one. This high
kurtosis and skewness in the far SOL region is mainly due to the frequent
ELM events. The relatively lower values of kurtosis and skewness of
the SOL QCF imply its non-intermittent features.

\subsection{Enhanced boundary transport by SOL QCF}

\begin{figure}
\includegraphics[width=3.3in]{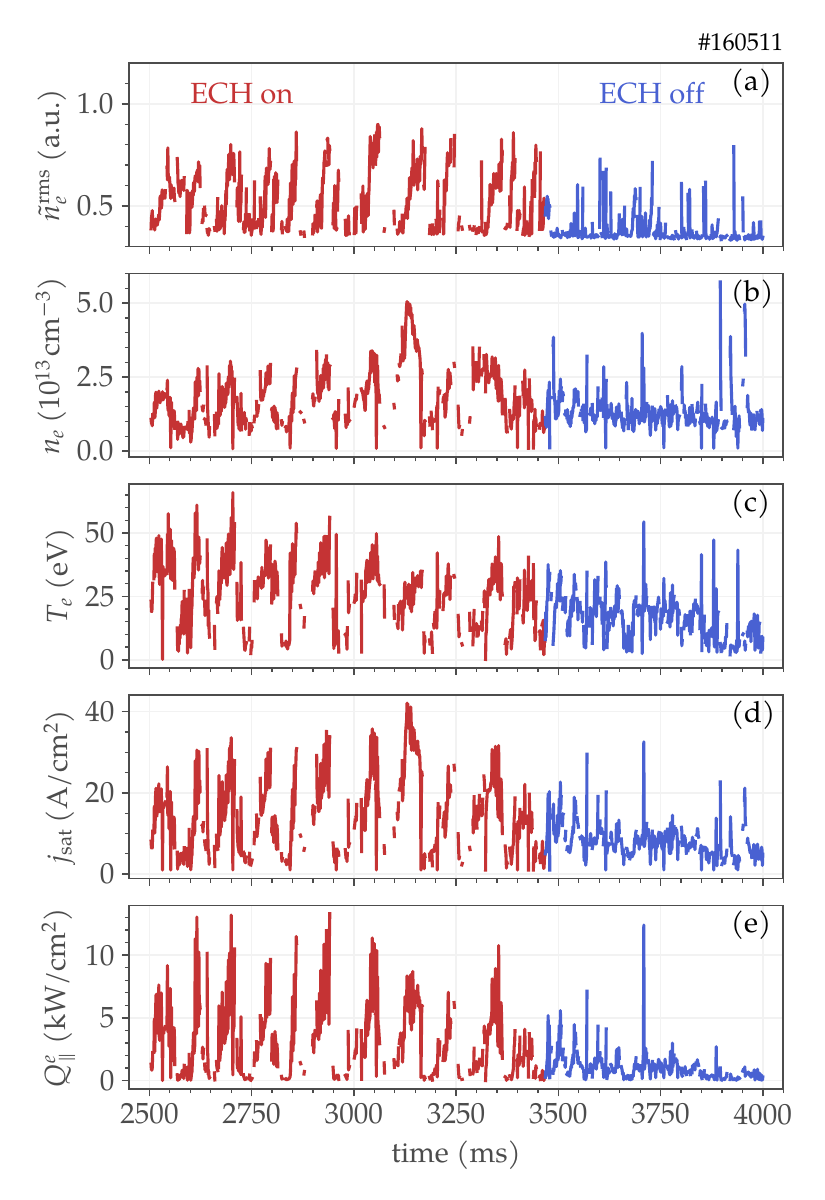}

\caption{\label{fig:probe} Time traces of the amplitude of the SOL QCF measured
by DBS (a), divertor electron density (b), divertor electron temperature
(c), the ion saturation current density (d), and the parallel heat
flux(e) onto the target in the `primary\textquoteright{} (upper) divertor,
respectively. Here, the divertor parameters are measured by fixed
divertor probes. Red and blue curves correspond to the time with and
without the ECH.}
\end{figure}

We also explored the potential effects of the SOL QCF on particle
and heat transport into the divertors. The fixed Langmuir probes \citep{watkins2008highheat}
are employed to provide the plasma parameters on the divertor targets,
such as electron density and temperature as well as the particle and
heat fluxes. These divertor quantities are then used to compare against
the midplane density fluctuation corresponding to the SOL QCF. Here,
we use the ion saturation current density, $j_{\mathrm{sat}}\propto n_{e}c_{s}$
with $c_{s}$ the ion sound speed, to indicate the particle flux into
the sheath of the divertor targets. In this study, the responses to
the large type-I ELMs in all signals are removed. For a typical discharge
shown in Fig.~\ref{fig:probe}, the SOL QCF is correlated with elevated
divertor electron density and temperature as well as the parallel
particle and heat fluxes. During the burst of the SOL QCF, there is
a clear increase in the electron density and temperature measured
near the strike points in the `primary' divertor (red in Fig.~\ref{fig:probe}(b)
and (c)). Also, the ion saturation current density and the parallel
heat flux increases during the burst of the SOL QCF (red in Fig.~\ref{fig:probe}(d)
and (e)). When the SOL QCF is suppressed, the divertor density, temperature,
and corresponding particle and heat fluxes are also reduced (blue
in Fig.~\ref{fig:probe}(b)--(e)).

\begin{figure}
\includegraphics[width=3.3in]{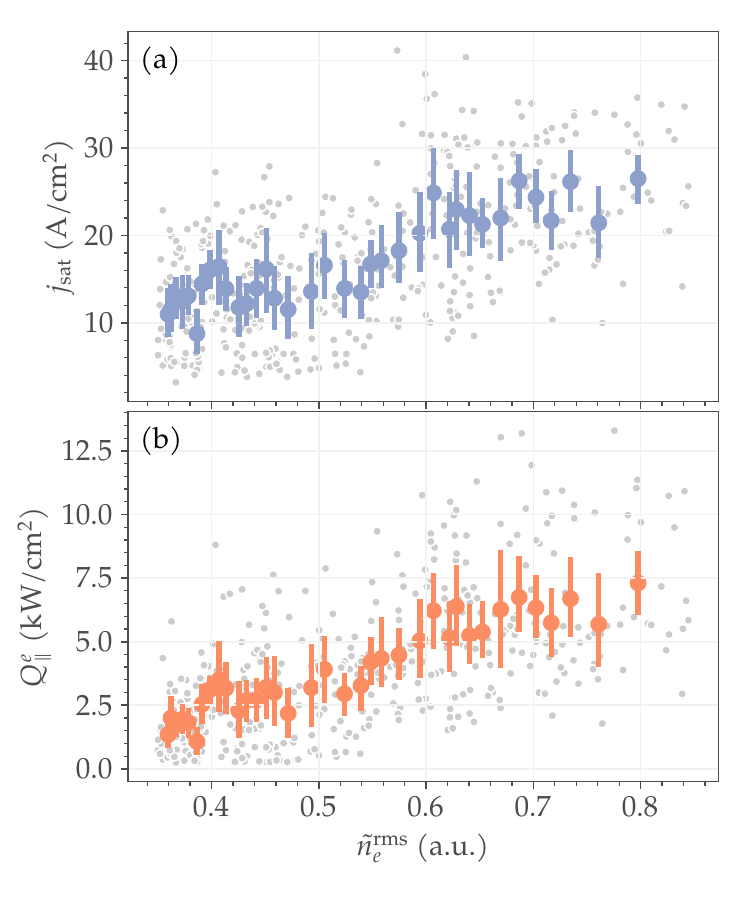}

\caption{\label{fig:fluxes}The ion saturation current density (a) and the
parallel heat flux (b) plotted against the density fluctuation level
corresponding to the SOL QCF. Errorbars show the mean and standard
deviation at each density fluctuation level. The underlying gray dots
represent the raw data. Both fluxes show clear increasing tendency
during the burst of the SOL QCF. }
\end{figure}

The ion saturation current density and the parallel heat flux are
plotted against the midplane density fluctuation level, to see the
how the divertor fluxes responds to the SOL QCF at the midplane. As
shown in Fig.~\ref{fig:fluxes}, the magnitude of the ion saturation
current density, $j_{\mathrm{sat}}$, and the parallel heat flux,
$Q_{\parallel}^{e}$, is greater during the burst of SOL QCF (e.g.,
$\tilde{n}_{e}^{\mathrm{rms}}>0.6$). In other words, the SOL QCF
can potentially increase the particle and heat fluxes onto the divertor
targets by a factor of 2--3. These findings imply the important role
of the SOL QCF in the boundary transport processes.

In these discharges, the coverage of the divertor Langmuir probes
was not optimized for this plasma shape, so detailed measurements
of particle and heat distributions on the divertor targets were not
available. Further studies are needed to investigate the effects of
the SOL QCF on the profiles of density and temperature, as well as
particle and heat fluxes, on the divertor targets.

\subsection{Influence of boundary profiles}

\begin{figure*}
\includegraphics[width=1\textwidth]{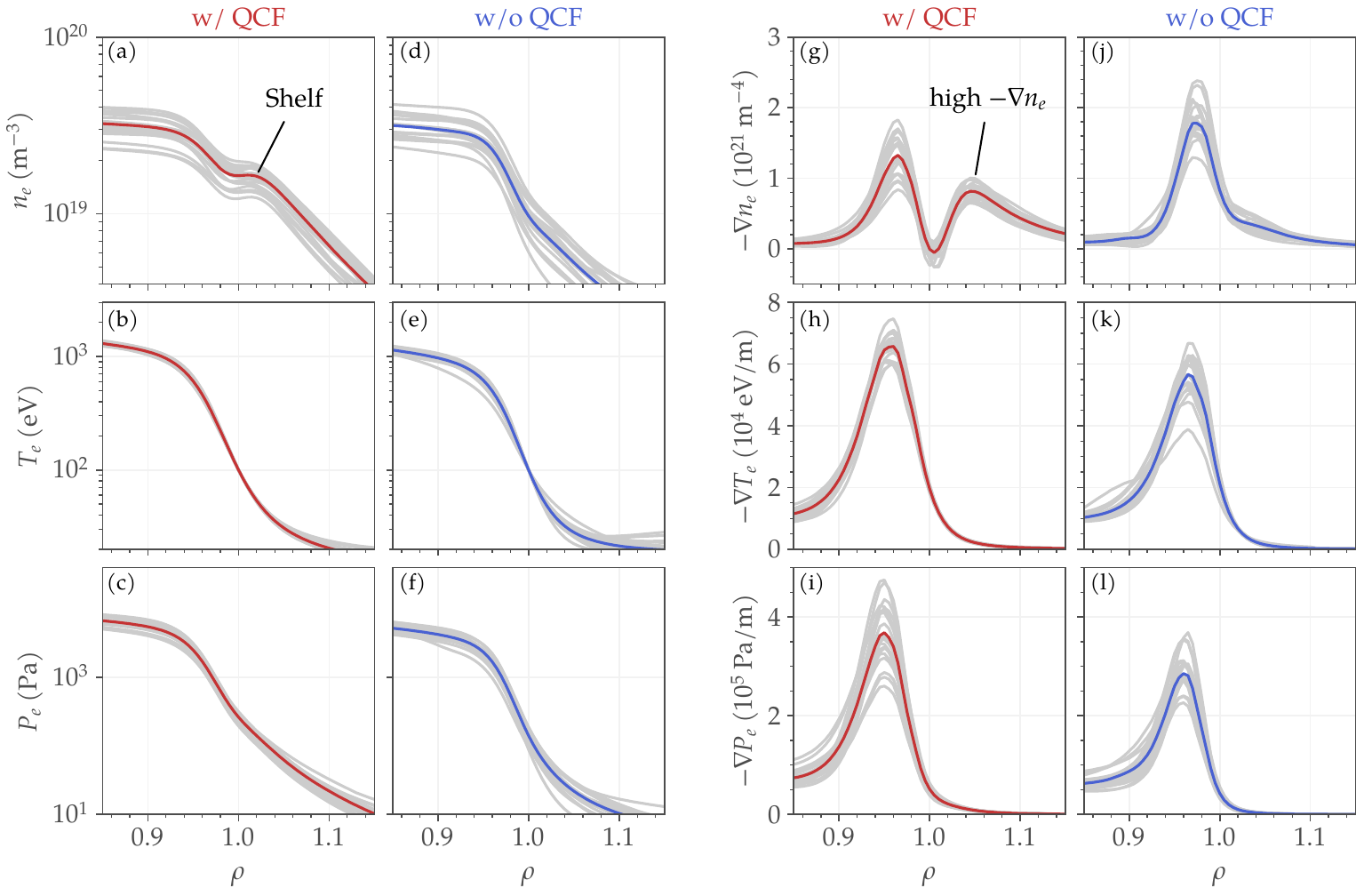}

\caption{\label{fig:profiles} Left panel (a)--(f): boundary profiles of the
electron density, temperature, and pressure. Right panel (g)--(l):
radial gradients of the electron density, temperature, and pressure.
Log-scales are applied in (a)--(f). Red and blue curves correspond
to mean profiles with and without SOL QCF, respectively. A shelf-like
profile can be identified near the separatrix and is associated with
higher SOL density gradient.}
\end{figure*}

The boundary plasma profiles and their links to the SOL QCF are examined.
We collected boundary profiles of electron density, temperature and
pressure from 20 shots of the same run-day. The boundary profiles
are conditionally averaged for time slices with and without SOL QCF
for each shot. Note that most of cases with the QCF observed occur
during the ECH phase. The profiles from each shot are then plotted
as gray curves in Fig.~\ref{fig:profiles}(a)--(f), where the cases
with QCF are plotted on the left-hand side and those without SOL QCF
are on the right-hand side. Averaging over all 20 shots gives the
mean profiles of the electron density, temperature, and pressure (red
and blue in Fig~\ref{fig:profiles}(a)--(f)). The red and blue curves
in each plot represent the mean profiles with and without SOL QCF,
respectively. In addition, the corresponding radial gradients with
and without SOL QCF are calculated (Fig~\ref{fig:profiles}(g)--(l)).
Similarly, the conditionally averaged radial gradients for each shot
are plotted in gray curves, and the shot-wise average of gradients
are plotted in red and blue for the cases with and without SOL QCF,
respectively. As shown in Fig.~\ref{fig:profiles}(a), during the
ECH phase, with the SOL QCF observed, the electron density profile
exhibits a shelf-like structure that spans the separatrix and extends
into the SOL region. This particular density profile leads to an enhanced
SOL density gradient (Fig.~\ref{fig:profiles}(g)).

It is worth noting that this density shelf structure across the separatrix
differs from the conventional SOL density shoulder. While the conventional
density shoulder is observed in high collisionality plasmas (typical
$\nu_{e}^{*}>1)$, the density `shelf' in DIII-D is only observed
in low collisionality plasmas (typical $\nu_{e}^{*}<0.5$) \citep{petrie2017improved,wang2020firstevidence}.
The formation of the density shelf might be attributed to the enhanced
SOL $E\times B$ drift flows that propagate from divertors to the
midplane and the resultant particle accumulation in the mainchamber
\citep{wang2020firstevidence}.

\begin{figure}
\includegraphics[width=3.3in]{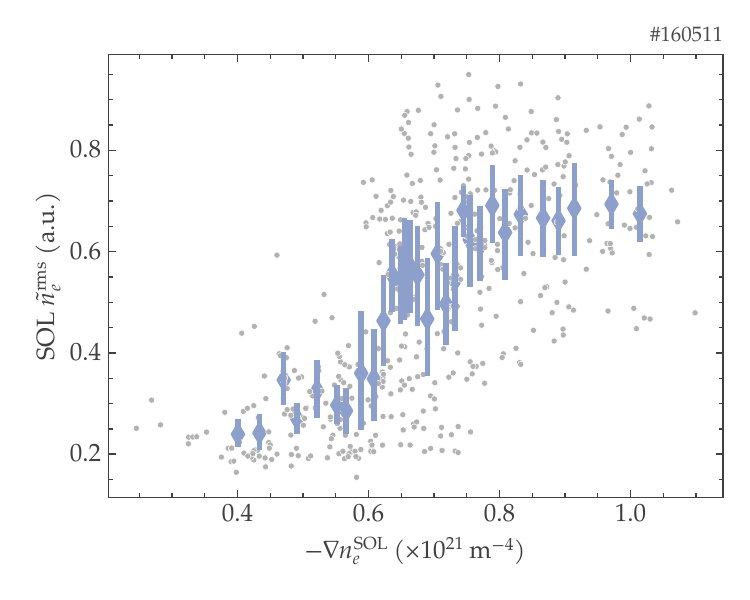}

\caption{\label{fig:sol-refl}The response of the SOL density fluctuation levels
in the QCF frequency range ($1.5\text{–}2.5$ MHz in DBS signals)
to the SOL density gradient (by reflectometry). The purple errorbars
show the mean and standard deviation of the density fluctuation level
at each SOL density gradient. The underlying gray dots indicate the
raw data.}
\end{figure}

According to the profile analysis above, the SOL density gradient
shows larger variation than other boundary gradients, and thus might
serve as the free energy drive for the SOL QCF. To investigate the
response of the SOL QCF to the SOL density profile, the amplitude
of the SOL QCF is compared against the SOL density gradient. Here,
the reflectometry diagnostics \citep{zeng2014performance} are employed
to provide the measurements of the density profile with higher time
resolution ($\delta t=1$~ms), so that we can trace the response
in each inter-ELM cycle. As shown in Fig.~\ref{fig:sol-refl}, the
higher density fluctuation levels of the SOL QCF are associated with
higher SOL density gradients. It exhibits a critical density gradient
behavior---the density fluctuation shows no clear response at lower
density gradient, but increases very rapidly once the critical gradient
is exceeded. These findings imply that $-\nabla n_{e}^{\mathrm{SOL}}$
plays an important role in the formation of the SOL QCF.

\begin{figure}
\includegraphics[width=3.3in]{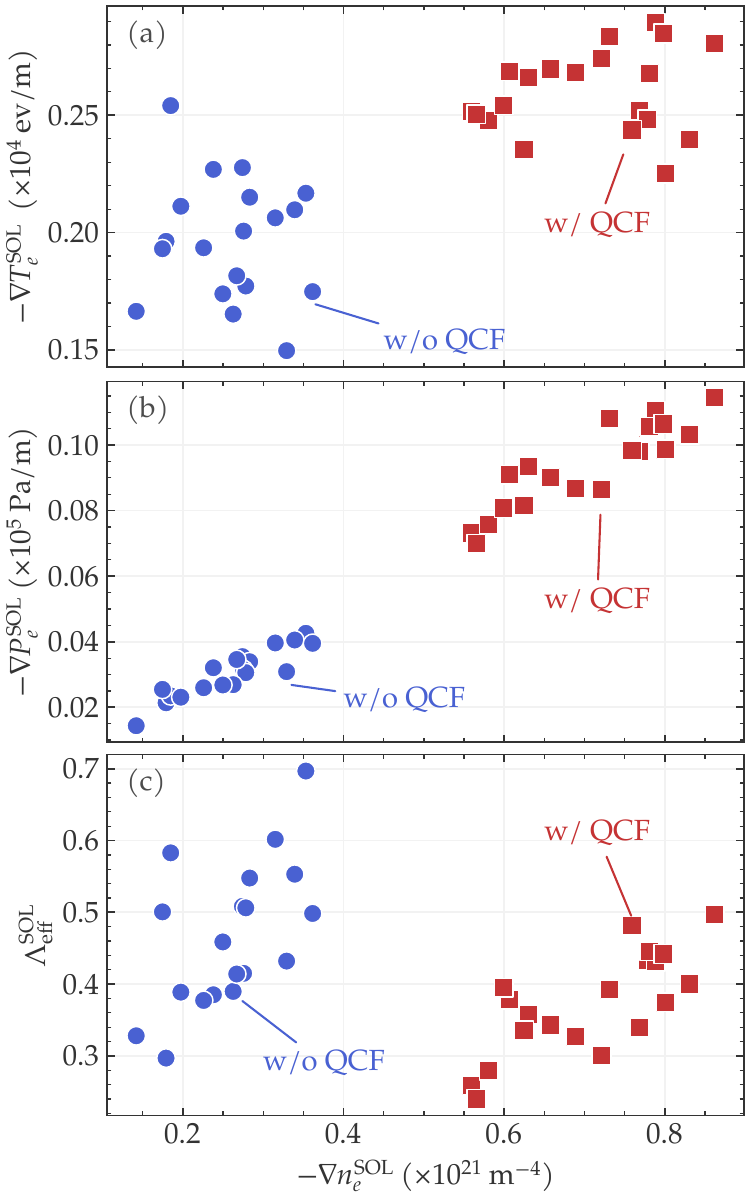}

\caption{\label{fig:sol-grad}The SOL temperature gradient (a), pressure gradient
(b), and the effective collisionality (c) plotted against the SOL
density gradient. The parameters are averaged over $1.01<\rho<1.08$.
Red squares and blue circles correspond to cases with and without
SOL QCFs, respectively.}
\end{figure}

The more general responses of the SOL QCF to the SOL electron density,
temperature, and pressure gradients are shown in Fig.~\ref{fig:sol-grad}(a)
and (b). The red and blue data points represent the cases with and
without SOL QCF observed, respectively. Here, the SOL electron temperature
gradient ($-\nabla T_{e}^{\mathrm{SOL}}$) and pressure gradient ($-\nabla P_{e}^{\mathrm{SOL}}$)
are plotted against SOL density gradients ($-\nabla n_{e}^{\mathrm{SOL}}$).
The gradients are averaged over $1.01<\rho<1.08$. Clearly, the SOL
QCF is found to be correlated with higher SOL density gradient (Fig.~\ref{fig:sol-grad}(a)).
It may also be associated with higher SOL electron temperature gradient
(Fig.~\ref{fig:sol-grad}(a)). However, the relative variation of
the SOL electron temperature gradient is much smaller than that of
the SOL density gradients. Specifically, the maximum to minimum ratio
of $-\nabla T_{e}^{\mathrm{SOL}}$ is about 2, while the ratio of
$-\nabla n_{e}^{\mathrm{SOL}}$ is about 5. As a result, $-\nabla P_{e}^{\mathrm{SOL}}$
is almost linearly proportional to $-\nabla n_{e}^{\mathrm{SOL}}$,
and the excitation of the SOL QCF is also associated with higher $-\nabla P_{e}^{\mathrm{SOL}}$
values (Fig.~\ref{fig:sol-grad}(b). These findings suggest that
either $-\nabla n_{e}^{\mathrm{SOL}}$ or $-\nabla P_{e}^{\mathrm{SOL}}$
can serve as the free energy drive for the SOL QCF.

The SOL and divertor collisionality has been previously observed to
have an influence on the dynamics of the SOL turbulence such as the
blobs \citep{carralero2015experimental}, and thus may also affect
the response of the SOL QCF. The effective SOL collisionality, $\Lambda_{\mathrm{eff}}^{\mathrm{SOL}}=\nu_{ei}L_{\parallel}\Omega_{i}/(\Omega_{e}c_{s})$,
is used to characterize the electron-ion collisional dynamics in the
SOL region \citep{myra2006blobbirth,carralero2015experimental}. Here,
$\nu_{ei}$ is the electron-ion collision rate, $L_{\parallel}$ is
the connection length, $c_{s}$ is the sound speed, and $\Omega_{e,i}$
is the electron/ion gyro-frequency. However, there is no obvious threshold
in the $\Lambda_{\mathrm{eff}}^{\mathrm{SOL}}$ that distinguishes
the cases with or without the SOL QCF. This finding indicates that
the excitation of the SOL QCF does not depend directly on the collisional
effects (Fig.~\ref{fig:sol-grad}(c)), although it is observed in
the low collisionality plasmas.

\subsection{Linear simulation using BOUT++}

A linear simulation using the BOUT++ framework with a reduced 5-field
fluid model has been carried out and compared to the observations
shown in previous sections. This model evolves five perturbed variables,
i.e., ion density, ion and electron temperature, magnetic flux, and
the vorticity \citep{li2019prediction,xia2015nonlinear}. This reduced
5-field model includes the ideal and resistive peeling-ballooning
physics, as well as diamagnetic and $E\times B$ drift. The simulated
domain in normalized poloidal flux is $0.95<\psi_{n}<1.10$, and the
grid resolution is $N_{\psi_{n}}=256$ in the radial direction and
$N_{y}=64$ in the poloidal direction. The resistivity $\eta$ is
considered as Spitzer-Härm resistivity. Since it is based on the Braginskii
reduced fluid equations, the model is only suitable for the simulation
of long wavelength turbulence with $k_{\bot}\rho_{s}<1$. This wavenumber
range covers the scales of the SOL QCF ($k_{\bot}\rho_{s}=0.2-0.4$).
BOUT++ employs only the toroidal mode number, $n$, as the independent
variable for the linear scan, and the poloidal mode number, $m$,
is determined via a local effective safety factor $q_{\mathrm{eff}}$
inferred from the local pitch angle \citep{dudson2009bouta}, i.e.,
$m=nq_{\mathrm{eff}}$. Thus, the effective poloidal wavenumber can
be defined as $k_{\theta}^{\mathrm{eff}}=nq/a$, where $a$ is the
plasma minor radius, $m$ and $n$ are the poloidal and toroidal mode
number respectively. More details about the BOUT++ 5-field model deployed
in this study can be found in previous publications \citep{li2019prediction,xia2015nonlinear}.

\begin{figure}
\includegraphics[width=3.3in]{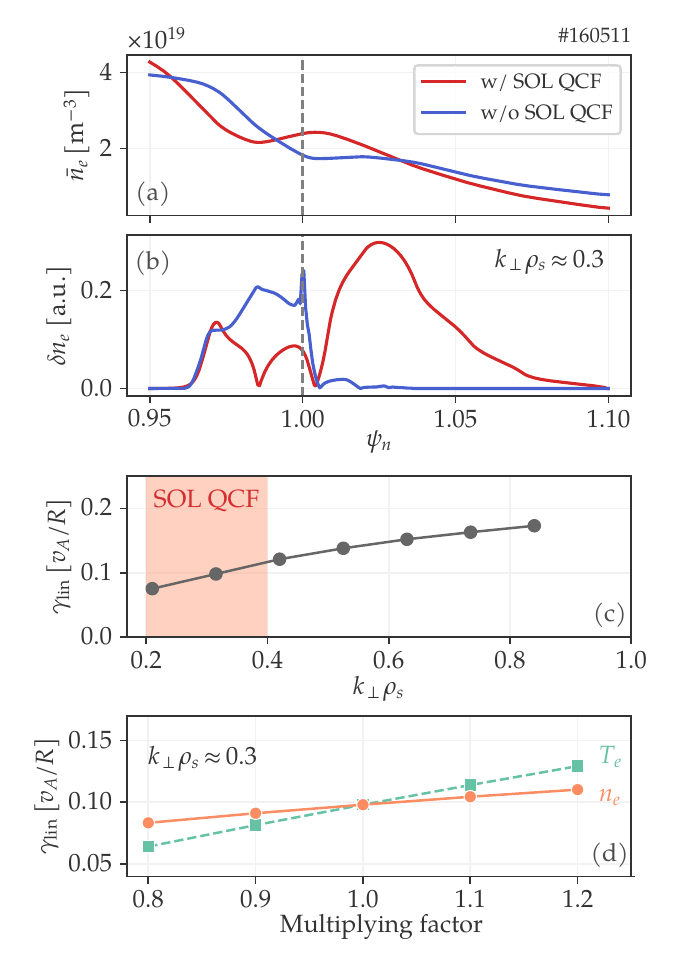}

\caption{\label{fig:bout}(a) Input density profiles with (red) and without
(blue) SOL QCF observed. (b) Density perturbation profiles with (red)
and without (blue) SOL QCF observed with $k_{\perp}\rho_{s}\approx0.3$.
(c) Linear growth rates calculated with experimental profile (red
in (a)) at different wavenumber; the shaded area indicates the wavenumber
range of the SOL QCF. (d) Linear growth rates calculated with different
scale factors of the electron density (orange) and temperature (green)
profiles. }
\end{figure}

In the linear BOUT++ simulations, experimental profiles at two different
time slices are used as the input for a comparative study. One set
of profiles is from 3250 ms of shot 160511 (with SOL QCF, red in Fig.~\ref{fig:bout}(a)),
and the other is from about 3850 ms (without SOL QCF, blue in Fig.~\ref{fig:bout}(a)).
When the SOL QCF is observed, the density gradient increases significantly
in the SOL region (red in Fig.~\ref{fig:bout}(a)). One may also
note that, in the latter case (blue in Fig.~\ref{fig:bout}(a)),
the SOL density profile is more flattened and looks more close to
conventional density shoulder.

The simulation results show that, with the experimental profiles at
3250 ms (red in Fig.~\ref{fig:bout}(a)), the density perturbation
can be destabilized in the SOL region (red in Fig.~\ref{fig:bout}(b)).
According to the BOUT++ linear simulations at $k_{\perp}\rho_{s}\approx0.3$,
the density perturbation, $\delta n=\tilde{n}/\bar{n}$, peaks in
the SOL region (red in Fig.~\ref{fig:bout}(b)). The peak of this
density perturbation is located at $\psi_{n}\approx1.02\text{–}1.03$,
corresponding to the location with maximum density gradient in the
SOL region. The radial width (FWHM) of this density perturbation is
about 1.3 cm. On the other hand, with the experimental profiles at
about 3850 ms (blue in Fig.~\ref{fig:bout}(a)), the SOL density
perturbation is stabilized (blue in Fig.~\ref{fig:bout}(b)). Also,
the pedestal density perturbation ($0.95<\psi_{n}<1$) becomes the
dominant mode when the SOL QCF does not appear (blue in Fig.~\ref{fig:bout}(b)).
The simulated mode is most unstable at the outer midplane, bad curvature
region, and propagates in the ion diamagnetic drift direction, indicating
its interchange-like nature. Above all, the propagation direction,
the peak location, and the radial width of the SOL density perturbation
are in agreement with the experimental observations by the DBS system.

The linear growth rate of the SOL density perturbation are calculated
at different turbulence scales ($0.2<k_{\perp}\rho_{s}<0.9$), using
the experimental profiles with SOL QCF observed. The SOL density perturbation
is found to be unstable for a wide range of the poloidal wavenumbers
which covers the range of the SOL QCF (Fig.~\ref{fig:bout}(c)). 

To identify the dominant free energy drive for the SOL density perturbation,
the density and temperature profiles are multiplied by a factor, ranging
from 0.8 to 1.2, with $k_{\perp}\rho_{s}\approx0.3$ in the simulation.
It is a common scanning technique used in BOUT++ simulations \citep{li2019prediction}.
The local quantity and gradient are thus changed accordingly, while
the scale lengths are fixed. As shown in Fig.~\ref{fig:bout}(c),
the linear growth rate increases as the density (orange) or the temperature
(green) profiles are raised. Since the increased density and temperature
profiles can enhance the linear growth rate of the SOL density perturbation,
its free energy source is likely attributed to the SOL pressure gradient.

\section{Discussion\label{sec:Discussion}}

In this study, we have identified the quasi-coherent fluctuation as
an important turbulence structure that may influence the SOL turbulent
transport in the high-power, high-performance hybrid plasmas. The
SOL QCF is correlated with the steepened SOL density and pressure
gradients, due to a shelf-like density profile across the separatrix.
Also, this QCF is an ion-scale long-wavelength fluctuation, and propagates
in the ion diamagnetic drift direction in the plasma frame. The linear
simulation using BOUT++ shows that the SOL density perturbation is
most unstable at the outer midplane bad curvature region, and the
SOL pressure gradient can provide the free energy drive for this mode.
These findings indicate that the SOL QCF is an interchange-like mode.

Conventionally, the blobs have been considered the dominant turbulence
structure in the SOL region \citep{boedo2009edgeturbulence}. Some
differences between the SOL QCF and blobs, as well as the implications
to the SOL transport studies, can be drawn: 
\begin{itemize}
\item The SOL QCF is a long-lived mode during the inter-ELMs period, while
the blobs are intermittent events. This is also reflected in the probability
density function (PDF): the the PDF of the SOL QCF is symmetrical
and has a small skewness value, while the blobs' PDF is featured
by a substantial skewness in the SOL region \citep{boedo2003transport}.
Thus, the SOL QCF may result in more contribution to SOL transport
than the blobs over a long period of time, which might be an issue
for long-pulse and steady-state operations. 
\item The SOL QCF shows little dependence on the effective SOL collisionality,
$\Lambda_{\mathrm{eff}}^{\mathrm{SOL}}$, while the blobs' size and
the induced transport is found to increase significantly as $\Lambda_{\mathrm{eff}}^{\mathrm{SOL}}$
is raised \citep{carralero2015experimental,russell2007collisionality}.
It is therefore expected that in low-collisionality plasmas the blobs
and its induced transport would be reduced, while the SOL QCF would
be less affected. The collisionality of plasma in future devices will
probably be much lower than that in present devices. Hence, the control
of the SOL QCF should be taken in to account in future fusion devices. 
\item The SOL QCF is a radially localized mode with a radial correlation
length of $l_{c,r}\approx1.5\text{–2}$ cm, while the blobs are
coherent objects that often propagate radially for a long distance.
So, the SOL QCF may be less likely to hit the mainchember wall and
thus have less impact than the blobs on the recycling and plasma-material
interaction in the mainchamber. 
\end{itemize}
Due to the observations and implications discussed above, the SOL
QCF might become an issue for the control of the plasma-material interaction
(PMI) in future fusion devices operated with high-power high-$\beta_{\mathrm{N}}$
scenarios. The observation of the SOL QCF may also raise a challenge
for developing more accurate transport and PMI modeling in the boundary
region of high-$\beta_{\mathrm{N}}$ plasmas. Clearly, further research
is needed to provide more detailed measurements of the SOL QCF and
its influence on SOL transport, including its effects on the particle
and heat deposition on the divertor targets and the mainchamber wall.

\section{Summary\label{sec:Summary}}

To summarize, a quasi-coherent density fluctuation (QCF) has been
observed in the SOL region of high-power, high-performance plasmas
with ECH injection and near double-null divertor configuration on
the DIII-D tokamak. The SOL QCF appears rapidly after the preceding
ELM within 10 ms. This mode is correlated with substantially enhanced
particle and heat fluxes (e.g., by a factor of 2--3) onto the divertor
targets, indicating its potential effects on the SOL transport in
high-$\beta_{\mathrm{N}}$ plasmas. The SOL QCF is associated with
higher SOL electron density and pressure gradients due to a shelf-like
density profile across the separatrix. The QCF propagates in the ion
diamagnetic drift direction $V_{\mathrm{ph}}\approx15\text{–20}$
km/s in the plasma frame and poloidal wavenumber $k_{\theta}\rho_{s}\approx0.2\text{–}0.4$.
The SOL QCF propagates outward in the radial direction with small
radial wavenumber $k_{r}\rho_{s}\approx0.03$, and the correlation
length is about $1.5\text{–}2$ cm. This SOL QCF does not show clear
dependence on the effective SOL collisionality $\Lambda_{\mathrm{eff}}^{\mathrm{SOL}}$.
These properties differ from those of the blobs that are conventionally
considered as the dominant SOL turbulence structures.

A linear simulation has been performed using BOUT++ with a 5-field
reduced model across the separatrix ($0.95<\psi_{n}<1.10$). The simulation
results confirm that a density perturbation can be driven by the elevated
SOL pressure gradient at the outer midplane bad curvature region.
The mode's propagation direction, peak location, and radial width
are in agreement with the experimental observations.

In conclusion, the SOL QCF is likely an interchange-like instability
driven by the SOL pressure gradients. It extends from the separatrix
to the far SOL region, and leads to significantly enhanced particle
and heat fluxes onto the divertor targets. The presence of the SOL
QCF may challenge present boundary modeling of high-$\beta_{\mathrm{N}}$
hybrid plasmas, and calls for more detailed work to evaluate its impact
on the SOL transport and plasma-material interactions.
\begin{acknowledgments}
The authors thank Dr.~B.~Grierson and Dr.~R.~Buttery for their
insightful comments on this paper. This material is based upon work
supported by the U.S. Department of Energy, Office of Science, Office
of Fusion Energy Sciences, using the DIII-D National Fusion Facility,
a DOE Office of Science user facility, under Awards DE-FC02-04ER54698,
DE-SC0019352, DE-FG02-08ER54984, and DE-NA0003525.

\textbf{Disclaimer}: This report was prepared as an account of work
sponsored by an agency of the United States Government. Neither the
United States Government nor any agency thereof, nor any of their
employees, makes any warranty, express or implied, or assumes any
legal liability or responsibility for the accuracy, completeness,
or usefulness of any information, apparatus, product, or process disclosed,
or represents that its use would not infringe privately owned rights.
Reference herein to any specific commercial product, process, or service
by trade name, trademark, manufacturer, or otherwise does not necessarily
constitute or imply its endorsement, recommendation, or favoring by
the United States Government or any agency thereof. The views and
opinions of authors expressed herein do not necessarily state or reflect
those of the United States Government or any agency thereof. 
\end{acknowledgments}

\bibliographystyle{apsrev4-1}
\bibliography{sol_qcf}

\end{document}